# Multimedia-Video for Learning

By Chua Kah Hean
(Hwa Chong Institution)
Oh Ming Yeo
(Fuchun Secondary School)
Wee Loo Kang and Tan Ching
(Educational Technology Division, MOE)

## What

Multimedia engages an audience through a combination of text, audio, still images, animation, video, or interactivity-based content formats. Along this vein, free platforms have been seen to allow budding enthusiasts to create multimedia content. For example, Google sites (Wee, 2012b) offer creative opportunities in website development that enable text insertion, still image, video and animation embedding, along with audio and hyper-interactive links to simulations (Christian & Esquembre, 2012; Wee, 2013; Wee, Goh, & Chew, 2013; Wee, Goh, & Lim, 2013; Wee, Lee, Chew, Wong, & Tan, 2015).

This chapter focuses on the video aspect of multimedia, which can be positioned as a component to any effective self-paced on-line lesson that would be available anytime, anywhere via computer or mobile devices.

> **Technologies**
>
> Possible video software
>
> Free:
> CamStudio.org
> Screencast-O-Matic
> Wink
>
> Paid:
> Camstasia Studio
>
> Suggested Hosting Platforms
>
> Public
> http://www.youtube.com/
> http://vimeo.com
>
> Restricted to MOE teachers:
> http://emedia.moe.edu.sg/

The multimedia video approach aims to help users overcome barriers in creating engaging, effective and meaningful content (Barron & Darling-Hammond, 2008) for teaching and learning in an online environment.

## Why

Salman Khan[1], founder of the Khan Academy, once advocated, "Let's use video to reinvent education". This could be done by giving students short videos to watch at home and then allowing them to do their 'homework' in the classroom. In this way, the teacher would be able to provide immediate assistance when students encounter difficulties or are keen to extend their learning further. Anyone can access abundant video resources on YouTube, Vimeo and eMedia that suit their teaching and learning needs.

The use and creation of videos can deepen teachers' professionalism[2] as they become designers (Wee, 2012a, 2012c; Wee, Goh, & Lim, 2013) of video clips targeted to explain ideas and concepts. As students view videos at home, it frees up school curriculum time and allows teachers and students to engage in richer discussions during class time. To nurture our students as knowledge creators, teachers can also mentor students to create their own video to explain to other students their creative ideas and facilitate discourse in the process. Local competition platforms may serve to promote the technological creations of students as seen in the annually held School Digital Media Awards organized by MOE-ETD.

---

[1] Salman Khan. (2011). Let's use video to reinvent education TED Talks Podcast, from http://youtu.be/nTFEUsudhfs

[2] Teachers — The Heart of Quality Education Retrieved 20 October, 2010, from http://www.moe.gov.sg/media/press/2009/09/teachers-the-heart-of-quality.php

| |
|---|
| **Awards** SDMA (School Digital Media Awards) |
| **Self-Directed Learning**<br><br>Ownership of learning<br>- Students can create their own videos to explain certain concepts<br><br>Management and monitoring of own learning<br>- Students manage their time effectively to watch these video and engage in richer discussions in class<br><br>Extension of own learning<br>- Students explore further videos online to learn beyond the curriculum |
| **Collaborative Learning**<br><br>Effective group processes<br>- Students can interactively contribute their own ideas clearly and consider other points of view objectively and maturely on the video comments<br>- Students ask questions to clarify and offer constructive feedback<br><br>Individual and group accountability of learning<br>- Students participate in group discussions to deepen the learning, depending on one another for success |

## How

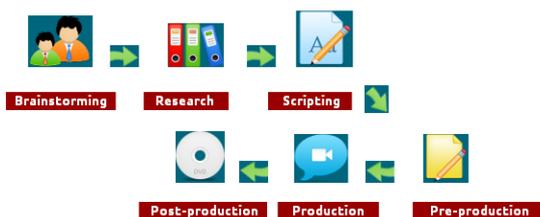

Figure 1. Video production guide-flow chart by MOE-ETD emedia with 6 steps of brainstorming, research, scripting, pre-production, prodcution and post production.

Video production comprises the key stages of pre-production (brainstorming, research and scripting), production (using the camera and video editing tools mentioned earlier), and post-production (able to quickly change the video after audience feedback). MOE-ETD emedia's 6 step brainstorming guide (0) has aptly outlined the respective stages in support of teachers and students' foray into the video creating process.

In the following section, we take a look at how one teacher used multimedia to enhance students' understanding of Physics concepts. Through this activity, students engaged in the process of scientific investigation using videos to illustrate concepts that would otherwise be a challenge to visualize (Wee & Goh, 2013) and learn.

| |
|---|
| **ETD Projects**<br>http://emedia.moe.edu.sg/<br><br>**Teacher-Led Video Projects**<br>Hwa Chong Institution<br>http://www.youtube.com/user/xmphysics/videos<br>Fuchun Secondary School<br>http://www.youtube.com/user/ohmingyeo/videos<br>Educational Technology Division<br>http://www.youtube.com/user/lookang/videos |
| **Resources**<br>Video Recording Device<br>Computers with Internet Connection<br>Video sharing platform |
| **Participation**<br>Join our Video Learning Community. Teachers will have access to lesson resources, community support, and consultancy.<br>For more information contact<br>Wee_Loo_Kang@moe.gov.sg<br>Tan_Ching@moe.gov.sg<br>chuakh@hci.edu.sg<br>oh_ming_yeo@moe.edu.sg |

# Lesson Plan

**DURATION:** 60 min          **SUBJECT:**     Physics

**LEVEL:** Upper Secondary to C 1          **TOPIC:**     Motion

*LEARNING OBJECTIVES*

*At end of the lesson, pupils should be able to:*

- Explain the physics involved in a given activity using video analysis

*MATERIALS AND RESOURCES REQUIRED FOR LESSON*

| ICT | Materials and Resources |
|---|---|
| Video Recording Device<br>Computers with Internet Connection<br>Video sharing platform | - |

*LESSON PROCEDURES AND PEDAGOGY*

| Time | Procedures | Pedagogies Used | Materials & Resources |
|---|---|---|---|
| 10 min | **Context**<br>This activity aims to take students through the process of a scientific investigation, using sports science, a real life example of long jump, billiard balls motion, butterfly swimming stroke, basketball throw, rotating fan.<br><br>At the start of the activity, teacher may provide any initial list of science demonstration and discuss steps involved in a scientific investigation. Teacher will then invite students to share previous scientific investigations carried out in primary science lessons. | Inquiry-based learning | http://weelookang.blogspot.sg/)<br><br>http://ictconnection.moe.edu.sg/ict-in-action&func=view&rid=727 |
| 20 min | Teachers are encouraged to introduce students to the pedagogy of video analysis and modeling using tracker if the motion is kinematics in nature. | Inquiry-based learning | |
| | Teacher may share that Science is not just a collection of facts or ideas about things around them. It is a way of thinking and finding out about the physical and natural world. This project-based lesson uses tools that professional physicists use for physics education.<br><br>Pose questions: This is the skill of asking suitable questions to initiate an investigation, in this case, a sport science that student find personally motivating.<br><br>Plan investigation: This is the process of devising ways to find the solution to the problem. This involves deciding on the measurements needed (displacement, velocity, acceleration, energies) and the types of equipment (video taken by the students, or otherwise) used to make these measurements.<br><br>Conduct an investigation: This is the process of carrying out the procedure to collect these measurements and observations. The measurements and observations are presented in a suitable way to report the findings to others.<br><br>Analyse and evaluate results: This is the skill of drawing conclusions from the investigation, and assessing whether the solution works, what is the physics being investigated.<br><br>Communicate results: This is the skill of presenting the conclusion to others in a coherent and logical manner.<br><br>Students are to first individually read through the problems from the other groups and then provide feedback to improve the latter's analyses.<br><br>**Note**: Close mentoring is required to guide student-groups through the entire process of investigation. | | |
| 20 min | Student-groups to present the scientific reports following a peer evaluation, contributing to a year-end formal assessment grade. | Collaborative peer learning | |

| 10 min | Teacher to mentor, grade and provide feedback on the areas of improvement for students to self-evaluate what can be improved on and for follow-up action by the students-groups.<br><br>Differentiating Instruction<br>For higher-ability students, teachers may guide the modeling construction to validate students' analysis and discuss the assumptions made in the investigation as well as provide suggestions for refinements to make the investigation better.<br><br>**Conclusion**<br><br>Teacher to emphasise that it is important to remain open-minded about the conclusions of the scientific investigation. Scientists should not be deterred if the outcomes of the investigation may not be what they have expected. Instead, one should reflect, evaluate and repeat the experiment again, if necessary. | Reflective thinking | |
| --- | --- | --- | --- |

*Lesson example from Mr Oh Jin sheng, Woodlands Ring Secondary School*

## Window into the Classroom

Mr Oh Ming Yeo teaches Physics at Fuchun Secondary School. As an avid user and creator of videos, he kindly shared some of his insights in teaching and learning Physics and the immense possibilities that multimedia could open up for students.

Firstly, there may be some weaker students who either do not fully understand certain concepts or require reinforcements on these concepts after classroom lessons.

Secondly, there are students who miss lessons due to illness or official duties. When faced with such problems, teachers will have to arrange for extra lessons after school. However, with the hectic schedule of a teacher, it is at times difficult to find time to help these students.

Thirdly, there is always a combination of stronger and weaker students in a class (Figure 2). There is therefore a need to cater to the varied pace of learning among students.

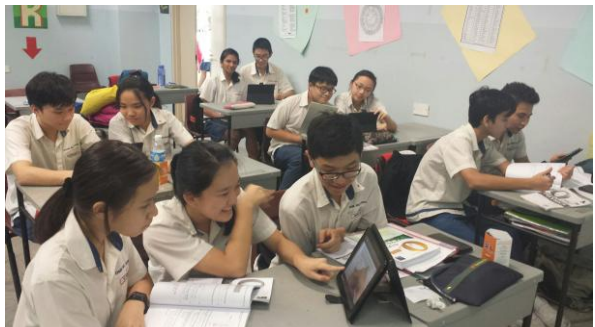

Figure 2. classroom in Fuchun Secondary where different ablity students progress at their own pace, meeting their own immediate learning needs. Photo from Oh Ming Yeo.

Lastly, weaker students who are unable to seek help in school may seek help from external tutors. However, not all students have the option of attending additional classes outside school.

To assist his students taking 'N', 'O' Science Physics, and 'O' Pure Physics, Ming Yeo created a series of 153 web-based Physics videos, which include topical concepts, experimental skills and 'O' Level detailed solutions for students to learn anytime.

Through the sharing of this resource and the feedback gathered on and offline, parents, students and teachers alike have reacted positively to how the channel has effectively enhanced students' learning capability. With just a smart phone connected to the Internet, topical revision can be done anytime, anywhere. The channel, set up one year ago has since been viewed more than 100,000 times locally and around the globe. There are currently 300 subscribers to the channel from all parts of the world.

Multimedia video also proved to be impactful for teaching and learning of advanced Physics concepts, even at the A-level. Mr Chua Kah Hean from Hwa Chong Institution recognized the effectiveness of media resource for both lectures and tutorials. The videos serve as good triggers for discussion and exploration of new concepts.

Kah Hean found that live demonstration is one way to engage learners but he also sees the advantages that recorded demonstrations can offer. He shared some useful examples below:

- Slow motion

Many physics phenomena happen so fast our brains cannot register the event. High speed video (Wee, Chew, Goh, Tan, & Lee, 2012) is very helpful in these situations.
AC light: http://youtu.be/SpkVFRJbBTo

- Annotations

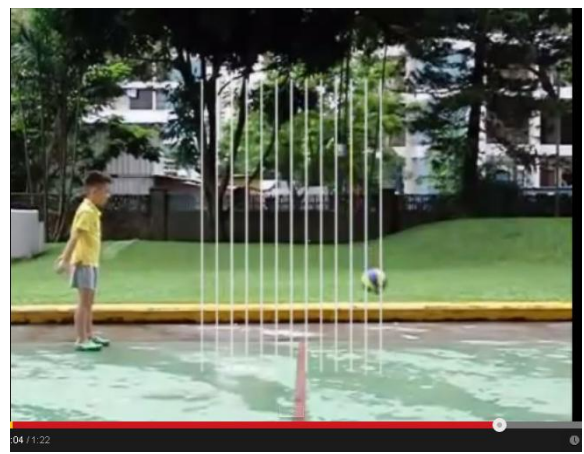

Figure 3. Projectile motion video showing equal horizontally spaced lines to illustrate uniform x direction velocity. YouTube by Chua Kah Hean

Sometimes, all it takes is a few lines added through video editing to turn an ordinary video into an insightful one.

Projectile motion (Figure 3):
http://youtu.be/Wxw2GY804t4

- Music

Some demonstrations are more enjoyable and impactful with background music.

Water fountain:
http://youtu.be/hSX3T6MjL9s

- Exploring Details

Many small size demonstrations are unsuitable for lectures because audience seated far away cannot see the necessary detail. Filming and playing the demonstration on the big screen solves the problem.

Tuning fork: http://youtu.be/LkORKcK6yVY

- Logistical Constraints

Videos allow demonstrations involving heavy and bulky equipment to be brought to classrooms easily.

Magnetic force: http://youtu.be/8J86xsjjMw8

As champions of using multimedia meaningfully in the classroom, both Ming Yeo and Kah Hean hope that the sharing of resources would invite more Physics teachers to come forward and contribute their own teaching videos (Wee, 2014) for a better web-based video resource bank that would benefit the entire Physics community(ISKME, 2008).